\documentclass[lettersize,journal,twocolumn]{IEEEtran}
\usepackage{graphicx}
\graphicspath{{Figures/}}
\usepackage{amsfonts}
\usepackage{amssymb}
\usepackage{amsmath}
\usepackage{color}
\usepackage{wasysym}
\usepackage{hyperref}
\usepackage{bbold}
\usepackage{algpseudocode}
\usepackage{placeins}
\usepackage{makecell}

\usepackage{mathtools}

\definecolor{lightblue}{RGB}{136,163,209}

\usepackage{algorithm}
\usepackage{algpseudocode}

\usepackage{authblk}

\begin{document}

\author[1]{Matteo Vandelli}
\affil[1]{Quantum Computing Solutions, Leonardo S.p.A., Via R. Pieragostini 80, Genova, Italy}
\author[1]{Francesco Ferrari}
\author[1,2]{Daniele Dragoni}
\affil[2]{Hypercomputing Continuum Unit, Leonardo S.p.A., Via R. Pieragostini 80, Genova, Italy}


\title{Parallel splitting method for \\ large-scale quadratic programs \\\vspace*{10pt} \normalsize  }

\maketitle



\begin{abstract}
Current algorithms for large-scale industrial optimization problems typically face a trade-off: they either require exponential time to reach optimal solutions, or employ problem-specific heuristics. To address these limitations, we introduce SPLIT, a general-purpose quantum-inspired framework for decomposing large-scale quadratic programs into smaller subproblems, which are solved in parallel. SPLIT heuristically accounts for objective-function cross-interactions between subproblems, usually neglected in other decomposition techniques. The SPLIT framework can integrate generic subproblem solvers, from branch-and-bound to quantum optimization methods. We demonstrate its effectiveness through comparisons with commercial solvers and published results on MaxCut and Antenna Placement Problems, with up to 20,000 variables. Our results show that SPLIT is capable of providing drastic reductions in computational time, while delivering high-quality solutions. In these regards, the proposed method is well-suited for near real-time applications that require a solution within a strict time frame, or when the problem size exceeds hardware limitations of dedicated devices, such as current quantum computers.
\end{abstract}

\section{Introduction}

Large-scale optimization problems in various industrial sectors---such as aerospace, energy, finance, healthcare, logistics, telecommunications, transportation---are often too complex to be solved exactly in a reasonable time due to the high number of decision variables~\cite{Petropoulos03032024}. In fact, problem-agnostic exact solvers, e.g. branch-and-bound~\cite{Land1960} and branch-and-cut~\cite{doi:10.1137/1033004} algorithms, typically require exponential time on classical digital computers as the problem size grows. Therefore, execution times can rapidly exceed practical computing capabilities, making exact solutions impractical for real-world applications. To overcome this issue, problem-specific heuristic algorithms have been devised to obtain near-optimal solutions efficiently~\cite{doi:10.1287/opre.21.2.498,10.1145/227683.227684}, prioritizing speed over solution quality. An important class of heuristics involves decomposition methods~\cite{doi:10.1287/opre.8.1.101,Benders1962,RePEc:eee:ejores:v:259:y:2017:i:3:p:801-817,conejo2006decomposition}, which break down large optimization problems into smaller subproblems, that can be solved separately with a much lower computational effort. Since combinatorial optimization problems involving integer, binary, or mixed-integer variables generally scale exponentially, their decomposition into smaller subproblems often results in a very large reduction of the computational time to obtain a near-optimal solution. The gain in computational efficiency is particularly valuable for industrial applications that require rapid solutions, e.g. in near real-time decision-making problems.

On a different note, recent advancements in quantum computing have sparked renewed interest in decomposition methods~\cite{https://doi.org/10.1002/qute.201900029,guerreschi2021solvingquadraticunconstrainedbinary,acharya2024decompositionpipelinelargescaleportfolio}, motivated by the fact that current quantum hardware cannot handle large-scale optimization problems due to size limitation. Decomposition pipelines have been integrated with various quantum algorithms to solve the resulting subproblems, such as quantum annealing~\cite{Rosenberg2016,booth2017partitioning} and the Quantum Approximate Optimization Algorithm (QAOA)~\cite{farhi2014quantumapproximateoptimizationalgorithm,zhao2025clusteringbasedsubquboextractionhybrid,Tomesh2022quantumlocalsearch,tomesh2023divideconquercombinatorialoptimization,chen2024noiseawaredistributedquantumapproximate}.

In this work, we propose a computational framework to tackle large-scale quadratic optimization problems which we name SPLIT, an acronym for Subproblem ParalleL Iterative Technique. This method is based on a decomposition of the whole problem into smaller subproblems, which are efficiently solved in parallel, exploiting distributed computing resources and/or high-performance computing (HPC) capabilities. 

In this context, the SPLIT method is a \emph{matheuristic} approach, as it leverages the specific structure of quadratic programs to guide the search for heuristic solutions. It is particularly well-suited for problems composed of weakly coupled blocks of variables, such as unconstrained binary problems~\cite{Kochenberger2014} with local connectivity or placement problems. Unlike other highly effective heuristic methods reported in the literature which are specific for unconstrained problems~\cite{LU20101254}, our approach can also accommodate certain types of constraints.

\begin{table*}[t]
\centering
\caption{Summary of notation used throughout Section~\ref{sec:math}.}
\label{tab:notation}
\begin{tabular}{ll}
\hline
\textbf{Symbol} & \textbf{Description} \\
\hline
$N$ & Total number of decision variables \\
$\mathcal{N} = \{1, \ldots, N\}$ & Index set of all decision variables \\
$x_i$ & $i$-th decision variable  \\
$X = \{x_i,\, i \in \mathcal{N}\}$ & Vector of all decision variables \\
$Q_{ij}$ & $(i,j)$-th coefficient of the QUBO matrix  \\
$H(X)$ & Quadratic cost function of the full problem  \\
\hline
$\mathcal{G} = (\mathcal{N}, E)$ & Undirected graph associated with the cost function \\
$E$ & Set of edges of graph $\mathcal{G}$ \\
\hline
$K$ & Number of subgraphs (i.e., subproblems)  \\
$\mathcal{G}_k = (S_k, E_k)$ & $k$-th subgraph \\
$S_k \subset \mathcal{N}$ & Set of nodes belonging to subgraph $k$ \\
$E_k$ & Set of internal edges of subgraph $k$ \\
$E_{kl}$ & Set of external edges connecting sites in different subgraphs $k\neq l$ \\
$X_k = \{x_i,\, i \in S_k\}$ & Decision variables assigned to subgraph $k$ \\
\hline
$d_i^{(k)}$ & Local field on node $i \in S_k$ due to variables outside $S_k$ \\
${D_k=\{d_i^{(k)}, \; i \in S_k\}}$ & Set of local fields for subgraph $k$ \\
$D = \bigcup_{k=1}^K D_k$ & Set of all local fields \\
$H_k(X_k, D_k)$ & Cost function of subproblem $k$ (including local fields) \\
$\Delta_k(X_k, D_k)$ & Double-counting correction term for subproblem $k$ \\
\hline
$N_{\rm iter}$ & Maximum number of iterations \\
$X^*$ & Heuristic solution returned by SPLIT \\
$X_{\rm min}$ & Exact optimal solution \\
\hline
\end{tabular}
\end{table*}

The solution is iteratively refined by a procedure that takes into account the cross-interactions between variables belonging to the different subproblems. From a computational standpoint, the overall scheme alternates (parallel) Jacobi-type and (sequential) Gauss–Seidel-type iterations~\cite{dyn1983numerical}. This shares similarities with block-coordinate descent approaches~\cite{warga1963,wright2015, doi:10.1137/120887679, doi:10.1137/16M1084705, Hou2016}, originally developed for convex optimization and continuous functions~\cite{Luo1992,Tseng2001,GRIPPO2000127}. On the other hand, SPLIT is directly applicable to combinatorial optimization problems with discrete variables, such as integer and mixed-integer quadratic programs, similarly to the methods discussed in~\cite{GONZALEZ20111826,Jaeger2020,ALABI20092472,Yang2024}. A key feature of the method is the explicit treatment of inter-subproblem couplings through the introduction of local fields, which embed the effect of external variables directly into each subproblem objective. This formulation is particularly relevant in the context of decomposition heuristics, where cross-interactions are often neglected or treated only indirectly~\cite{acharya2024decompositionpipelinelargescaleportfolio}. Furthermore, the use of spectral clustering to construct weakly coupled partitions plays an important role in enhancing the effectiveness of the approach by reducing inter-subproblem interactions from the outset.

The SPLIT framework can exploit any quadratic program solver for the solution of the subproblems. For instance, state-of-the-art commercial solvers, e.g. CPLEX~\cite{cplex2022v22} and Gurobi~\cite{gurobi}, can be integrated in SPLIT, further extending their range of applicability to much larger problem sizes, while keeping the computational time below a certain threshold. This is particularly relevant for real-time industrial applications, where obtaining a solution in a short time is necessary. Additionally, the decomposition to smaller subproblems within SPLIT enables the use of dedicated hardware, such as current quantum processing units (QPUs), which struggle with scalability and are limited by the size of the problems they can handle. In the case of constrained quadratic programs, SPLIT can incorporate certain types of constraints to ensure a feasible solution, provided that the subproblem solver is capable of enforcing hard constraints. Other kinds of constraints can be included as soft penalty terms.

This work is organized as follows. We first discuss the mathematical formulation behind the SPLIT framework and its implementation as a parallel algorithm. Afterwards, we benchmark the method on quadratic binary problems, using CPLEX as a subroutine for the solution of the subproblems. We provide a thorough comparison of SPLIT results with CPLEX applied to the full problem, focusing both on the solution quality and the required computational time. Our calculations involve Maxcut, a prototypical quadratic unconstrained binary optimization (QUBO) problem~\cite{karp1972}, and the Antenna Placement Problem (APP), which has potential application in the field of telecommunications~\cite{vandelli2024evaluating}. We show that a preliminary (not fully optimized) implementation of SPLIT already enables near-optimal solutions for large-scale problems with up to tens of thousands of variables on a single compute node. The possibility to exploit HPC resources through the inherent parallelism of SPLIT enables scalability towards even larger systems.

\section{Algorithm description}\label{sec:math}

\subsection{Mathematical framework}

Quadratic programs are optimization problems with a quadratic cost function to be minimized, possibly subject to a set of linear constraints. We defer the discussion of constraints to subsection~\ref{sec:constraints} and focus here on the quadratic cost function 
\begin{align}\label{eq:quadratic_problem}
    H(X) = \sum_{(i,j) \in E} x_i Q_{ij} x_j + \sum_{i\in \mathcal{N}} Q_{ii} x_i,
\end{align}
where the decision variables of the problem, denoted by $x_i$, can be either integer, continuous, or a combination of both. They are collected into the vector ${X = \{x_i, \, i \in \mathcal{N}\}}$, with $\mathcal{N} = \{1, ..., N\}$. The coefficients $Q_{ij}$ form the so-called QUBO matrix of the problem, which is assumed to be symmetric without loss of generality, i.e. $Q_{ij}=Q_{ji}$. We exploit the isomorphism between QUBO problems and graphs to introduce the associated (undirected) graph $\mathcal{G}=(\mathcal{N}, E)$, where the set $\mathcal{N}$ represents the graph \emph{nodes} and the set $E$ identifies the graph \emph{edges}, which are pairs of nodes $(i,j)$. This mapping is achieved straightforwardly by assigning diagonal elements $Q_{ii}$ to the nodes of the graphs, while off-diagonal elements $Q_{ij}$ are assigned to the edges. 

\begin{figure}[t]
    \centering
\includegraphics[width=0.65\columnwidth]{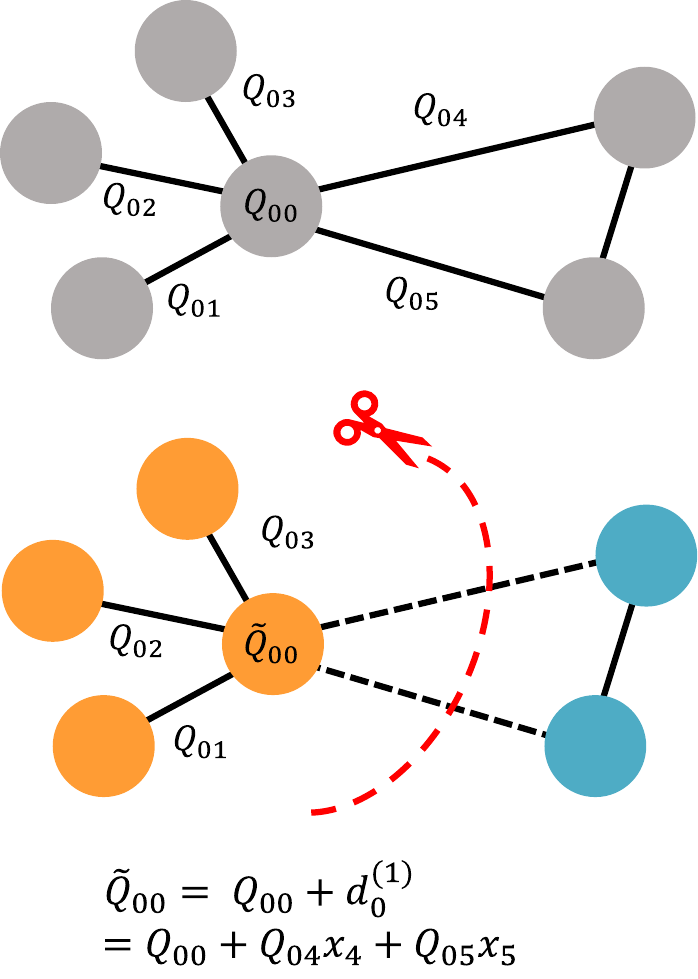}
    \caption{Example of graph partitioning. The original graph, depicted on the top with gray nodes, is partitioned into two subgraphs, indicated by the different colors on the bottom. The lower figure shows how the local field $d_0^{(1)}$ accounts for the external interactions acting on node $0$. The definition of Eq.~\eqref{eq:local_weights} preserves the local structure of the problem by evenly balancing the impact of internal and external edges.}
    \label{fig:split_graph}
\end{figure}

We begin by exactly partitioning the graph into $K$ smaller subgraphs $\mathcal{G}_k=(S_k,E_k)$, with $k \in \{1, ..., K\}$ being the subgraph index. The decision variables associated with the nodes of $S_k \subset \mathcal{N}$ form the set
\begin{align}
 X_k = \{x_i, \, i \in S_{k}\}. 
\end{align}
The edges of $\mathcal{G}_k$ are those which connect pairs of nodes within the subgraph, namely
\begin{align}
    E_k = \{(i,j) \in E|\, i \in S_k, j \in S_k \}.
\end{align}
In addition to these internal edges, external terms are present, connecting sites belonging to two different subgraphs
\begin{align}
    E_{kl} = \{(i,j) \in E|\; i \in S_k, j \in S_l , k \neq l\}.
\end{align}
Based on the above graph partitioning, we reformulate the cost function of the original quadratic problem~\eqref{eq:quadratic_problem} as a sum of the cost functions of the individual subgraphs
\begin{align}\label{eq:Hk_cost}
    H_k(X_k, D_k) = \sum_{(i,j) \in E_k} x_i Q_{ij} x_j + \sum_{i \in S_k} \left(Q_{ii} + d^{(k)}_i\right) x_i.
\end{align}
where the local weights are defined as
\begin{equation}\label{eq:local_weights}
d_i^{(k)} = \sum_{\substack{l=1 \\ l \neq k}}^K\sum_{(i,\alpha) \in E_{kl}} Q_{i\alpha} x_{\alpha}.   
\end{equation}
and form the sets ${D_k=\{d_i^{(k)}, \; i \in S_k\}}$. The latter are collected in the set $D=\bigcup_{k=1}^K D_k$. In practice, $H_k$ accounts for all the \textit{internal} interactions between nodes within the subgraph, plus the \textit{external} interactions with the nodes of other subgraphs through the $D_k$ fields. The presence of the $D$ terms introduces the dependence among the subproblems, which therefore cannot simply be treated as independent. 
Importantly, the definition of $d_i^{(k)}$ in Eq.~\eqref{eq:local_weights} ensures that both internal and external interactions are treated equivalently in the subgraph cost function $H_k$ (see Fig.~\ref{fig:split_graph}). However, this approach results in a double-counting of the external edges when reconstructing the global cost of the original problem by summing over all $H_k$. Therefore, we need to include correcting terms of the form
\begin{align}
    \Delta_k(X_k,D_k) = \frac{1}{2}\sum_{i \in S_k} d^{(k)}_i x_i
\end{align}
The problem can then be exactly reformulated in terms of a minimization problem over $x_i$ and $d_i^{(k)}$ 
\begin{align}
    \min_{X} H(X) \rightarrow  \min_{X,D} & \sum_{k=1}^K  \Big( H_k(X_k,D_k)  - \Delta_k(X_k,D_k) \Big) \notag\\
    & \text{subject to} \notag \\
    &d_i^{(k)} = \sum_{\substack{l=1 \\ l \neq k}}^K\sum_{(i,\alpha) \in E_{kl}} Q_{i\alpha} x_{\alpha} \notag\\
    & d_i^{(k)} \in {\mathbb{R}}, \; \; \forall \, i \in S_k, \; \;  \forall \, k \in \{1, ..., K\},
    \label{eq:approximations}
\end{align}
where we turned the $D$ local weights into decision variables by adding a trivial constraint on their values. 

The problem is decomposed since each term $H_k$ in the summation contains only variables $X_k$ and $D_k$. However, the subproblems are not separable since they are subject to the constraint of Eq.\eqref{eq:approximations}.
At this point, we observe that the local fields $D_k$ involved in subproblem $H_k$ are determined solely by the values of the $x_i$ decision variables belonging to the other subgraphs. If this dependence is weak, namely the intra-subproblem interactions are stronger than the inter-subproblem ones, i.e.
\begin{equation}
\sum_{\alpha \in \mathcal{N} \setminus S_k} |Q_{i\alpha}| \ll \sum_{j\in S_k} |Q_{ij}|, \; \; \forall i \in S_k, \,
\label{eq:block_diag_dominance}
\end{equation}
then the constrained minimization over the $D$ variables can be performed via greedy or local iterative methods, producing high-quality heuristic solutions to the original problem. 
The formulation of Eq.\eqref{eq:approximations} is the basis of the SPLIT method and allows us to devise a numerical strategy to solve each subproblem $H_k$ separately and in parallel, as described in the next paragraph.

\subsection{Implementation of SPLIT}

\begin{figure*}[!t]
    \centering
    \includegraphics[width=0.75\linewidth]{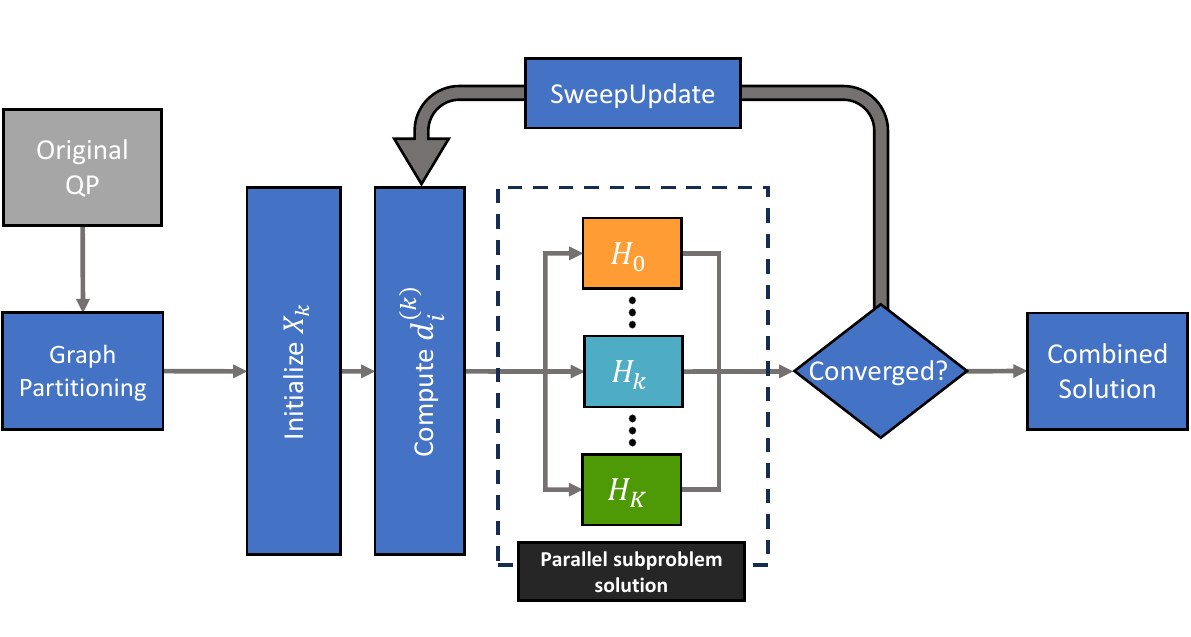}
    \caption{Schematic representation of the SPLIT framework for quadratic programs (QP). The flow diagram outlines the key stages, starting with the graph partitioning step, which determines the variables $X_k$ to be assigned to each subproblem, followed by initialization, iterative updates, and the final convergence check. Each step is designed to balance computational efficiency with solution accuracy.}
    \label{fig:algorithm_scheme}
\end{figure*}

Here, we present the SPLIT framework itself, which is the proposed practical way of solving Eq.\eqref{eq:approximations} approximately to obtain a near-optimal solution to the original problem. This methodology is designed to distribute the subproblems in parallel across multiple processing units, and uses an iterative procedure to update the $D$ variables of the minimization problem of Eq.\eqref{eq:approximations}. The pseudocode of the algorithm is shown in Algorithm~\ref{alg:par}, while a visual scheme is shown in Fig. \ref{fig:algorithm_scheme}. 

The method proceeds as follows. As a first step, we perform the partitioning of the full graph $\mathcal{G}$ using a clustering algorithm. Each node in the graph, corresponding to a variable \( x_i \), is then assigned to a particular subgraph $\mathcal{G}_k$. We adopt the spectral clustering method, which is well-suited for this task, as it groups nodes into clusters that maximize internal connection weights~\cite{868688, 1238361}. This approach facilitates the identification of partitions that, whenever possible, satisfy the condition in Eq.~\eqref{eq:block_diag_dominance}. The graph partitioning is performed only once, at the beginning of the calculation. 
The iterative procedure begins after the initialization of all decision variables $X$ to a set of predefined arbitrary values. Specifically, the choice $x_i=0, \ \forall i$ ensures that $D=0$ at the first iteration, which means that the various subproblems are initially decoupled. Alternatively, the process can begin from a different starting point derived from other approximation methods, e.g. continuous relaxation. 

Following initialization, three consecutive steps are executed and repeated iteratively until convergence is achieved. First, we compute the on-site fields $D$ as in Eq.~\eqref{eq:local_weights} and use them to set up the cost function of each subproblem $H_k$, following Eq.~\eqref{eq:Hk_cost}. Next, we independently solve each subproblem in parallel, finding the solutions $X_k$ that minimize the respective cost functions $H_k$, keeping the variables of $D_k$ frozen. The full cost function $H(X)$ is then evaluated at the aggregated vector $X=\bigcup_{k} X_k$. Finally, the last step of the iterative procedure is to check for convergence. If the newly computed cost function matches the value at the previous step, or the maximum number of step $N_{\rm iter}$ is reached, the procedure terminates and the heuristic solution $X^*$ is returned. Otherwise, a Sweep Update operation is applied to the decision variables $X$ and a new iterative step begins with the calculation of the updated local fields $D$. Here, Sweep Update generally refers to any greedy routine which runs over all the $X$ variables of the problem, sequentially proposing local updates involving a limited subset of variables, and accepting them whenever they reduce the global cost function. The simplest Sweep Update operation is to sequentially propose a single variable change $x_i$, as discussed below. More complex updates involving larger variable neighborhoods may improve performance, especially in constrained problems. In a broader context, SPLIT can be viewed as a generalized iterative procedure that alternates a Jacobi-type block coordinate minimization (the parallel subproblem solution) and a Gauss-Seidel-type minimization (Sweep Update)~\cite{wright2015}.

\begin{algorithm}[t]
\caption{SPLIT algorithm}\label{alg:par}
\begin{algorithmic}
\State {\bf Input }  $Q_{ij}$, $i,j=1, ..., N$ 
\State \phantom{{\bf Input }} {\rm number of clusters} $K$
\vspace{0.1cm}
\State $\mathcal{G} \gets {\rm \textsc{GenerateGraph}}(Q)$
\State $\mathcal{G}_k$ $\gets$ ${\rm \textsc{GraphPartitioning}}$($\mathcal{G}, {\rm num\_clusters} = K$)
\State $X \gets {0}$
\For{$n = 1, ..., N_{\rm iter}$} 
    \vspace{0.05cm}
    \State $D \gets$ \textsc{ComputeFields}($X$, $Q$)
    \vspace{0.05cm}
    \State {\bf{parallel for} $k=1, ..., K$ }
    \State \quad  $X_k \gets$ \textsc{Solve}($H_k$, local\_fields=$D_k$)
    \State {\bf{end parallel for}}
    \vspace{0.05cm}    
    \State $X \gets {\textsc{Concatenate}}(X_k)$
    \State $H_{\rm old} \gets H_{\rm new}$
    \State $H_{\rm new} \gets H(X)$
    \vspace{0.05cm}        
    \If{$H_{\rm new} = H_{\rm old}$}
    \State \quad $X^* \gets X$ 
    \State \quad {\bf return $X^*$}
    \EndIf
    \vspace{0.05cm}        
    \State $X \gets $ \textsc{SweepUpdate}($X$)
\EndFor
\end{algorithmic} 
\end{algorithm}

The core idea of this algorithm is to accurately capture the internal interactions within each subproblem, while approximating the external interactions using the \( D \) fields. As the iterative procedure progresses, we have empirically observed (Section~\ref{sec:comp_scaling}) that the \( D \) fields gradually converge to stable values.
The treatment of external interactions constitutes the main difference with respect to other decomposition methods, in which the subproblems are treated as fully independent, neglecting cross-interactions~\cite{acharya2024decompositionpipelinelargescaleportfolio}. The Sweep Update routine is introduced to scramble the variables in a way that is independent of the chosen graph partition, while improving the current solution cost. This helps reduce the risk of getting trapped in local minima, particularly in the early stages of the iterative procedure.

No general rule for selecting the number of subproblems $K$ was observed. The choice of $K$ is problem-specific and guided primarily by computational resources and runtime requirements. Still, the SPLIT method exhibits two trivial limiting cases. The first occurs at $K=1$, where the method yields the exact solution to the problem. The second limit, at $K=N$, reduces to performing greedy sweeps that locally update the variables—a simple yet often inaccurate heuristic. Intermediate cluster sizes $K$ provide a balance between these extremes, effectively interpolating between exact optimization and purely local updates.

The choice of the subproblem solver and the details of the Sweep Update routine are the only components of SPLIT that depend on the variable types (binary, continuous or mixed). Indeed, the decomposition steps leading to Eq.~\eqref{eq:approximations} do not rely on any assumption about the variables. In the present work, we focus on quadratic problems with binary decision variables $x_i \in \{0, 1\}$, although we stress that the proposed methodology can be applied to integer and mixed-integer problems, in general. For the binary problems tackled here, we adopt the most straightforward Sweep Update strategy: we sequentially iterate over all variables in a single sweep and for each variable we propose a flip. If the new configuration yields a lower cost, the flip is accepted; otherwise, it is discarded. This approach guarantees that each update does not increase the cost. The Sweep Update routine used in this work is summarized in Algorithm~\ref{alg:sweepupdate}.

\begin{algorithm}[t]
\caption{Sweep Update (for binary variables)}
\begin{algorithmic}
\State {\bf Input }  $Q_{ij}$, $i,j=1, ..., N$ 
\State \phantom{{\bf Input }} {\rm current solution} $X=\{x_1,\dots,x_N\}$
\vspace{0.1cm}
\For{$l = 1, ..., N$} 
    \State  $X^\prime \gets \{x_i + (1-2x_i)\delta_{i,l}\}_{i=1,\dots,N}$
    \If{$H(X^\prime) < H(X)$}
    \State \quad $X \gets X'$ 
    \EndIf
\EndFor
\State {\bf return $X$}
\end{algorithmic} \label{alg:sweepupdate}
\end{algorithm}

\subsection{Constraints}\label{sec:constraints}

In the formulation discussed so far, SPLIT serves as a heuristic approach for solving QUBO problems. In this subsection, we describe how the method can be extended to accommodate various constraints commonly encountered in industrial applications. A first approach is to incorporate them into the cost function as soft penalty terms, possibly introducing slack variables to manage inequality constraints. This approach is widely employed to impose constraints in numerical solvers that do not inherently support them, such as quantum annealing systems. A key drawback of this method is that it does not explicitly restrict the solver to explore only the space of feasible solutions.

However, certain specific constraints can be seamlessly integrated within SPLIT as hard constraints. When combined with solvers that enforce constraint satisfaction at the subproblem level, such as branch-and-bound, these constraints can be implemented in a way that ensures global feasibility of the final solution across the entire problem. Some examples of constraints that can be enforced exactly in SPLIT are:

\begin{itemize}
    \item \emph{Capacity constraints}, which can be expressed in the forms $\sum_i x_i = V$ or $\sum_i x_i \leq V$, with $V$ a real number. For instance, this global constraint can be included approximately by setting the constraints 
    \begin{align}
        \sum_{i \in S_k} x_i = \left\lfloor \frac{V |S_k|}{N}\right\rfloor + r_k
        \label{eq:cardinality_const}
    \end{align}
on each subproblem $H_k$, $k=1, ..., K$. Here, the $r_k$ terms account for the redistribution of the remainder ${ r = \left(V  - \sum_k \left\lfloor V \frac{|S_k|}{N}\right\rfloor\right)}$ according to an appropriate criterion. The fraction of $V$ assigned to each subproblem could in principle also be updated iteratively to improve the approximation.
    \item \emph{Local constraints} of the form
    \begin{align}
        \sum_{i \in S} A_i x_i \leq b \ \mbox{ or } \ 
        \sum_{i \in S} A_i x_i = b,
    \end{align}
    with $S$ a small subset of the decision variables of the whole problem. This can be dealt with by including all the variables appearing in the constraints in the same subproblem, so that $S \subseteq S_k$. If this is possible, then the constraint can be imposed on the $k$-th subproblem only without further processing.
\end{itemize}

Beyond enforcing constraints to ensure a feasible solution, it is also possible to introduce variations of the Sweep Update routine that preserve these constraints. This has the advantage of always improving the solution without leaving the feasible subspace, potentially leading to higher solution quality compared to a generic update. However, for general constraints, this is effectively impractical, as finding a feasible solution falls within the same complexity class as finding an optimal one. In some of the specific cases discussed above, this can be achieved with only a polynomial overhead. For instance, a possible update that preserves the global capacity constraint in the case of binary decision variables is the double bit-flip. This update involves identifying all pairs of variables $x_i$ and $x_j$ that belong to different subproblems and have opposite values in the current solution. The algorithm then evaluates whether flipping both bits would reduce the cost function. If the cost function decreases, the update is accepted; otherwise, the move is rejected.

That said, a constraint-preserving Sweep Update is not strictly necessary to maintain the feasibility of the solution if constraint satisfaction is enforced at the subproblem level. While the Sweep Update may temporarily violate constraints, potentially leading to configurations outside the feasible solution space, these violations are immediately corrected by the subproblem solver. By following the steps in the correct sequence, the SPLIT method always yields a feasible solution, provided the chosen subproblem solver respects the constraints.

\section{Performance metrics and computational details}

We assess the effectiveness of SPLIT by several performance metrics. For this purpose, we choose two different problems, the MaxCut problem and the APP, both of which can be formulated as quadratic optimization problems with binary decision variables.
MaxCut is a notoriously NP-hard problem on general instances and is unconstrained, while the APP has a capacity constraint on the number of antennas that can be turned on simultaneously. 

We present a comparative analysis of three distinct methods:
\begin{itemize}
\item the SPLIT method introduced in this work, using the CPLEX commercial solver by IBM~\cite{cplex2022v22} as subproblem solver. 
\item CPLEX Exact: This method directly utilizes multi-threaded CPLEX for the whole problem, without imposing a maximum duration for the optimization process. CPLEX is run using its default MIP termination criteria, with relative and absolute optimality gap tolerances of $10^{-4}$ and $10^{-6}$, respectively. Under these settings, CPLEX terminates once the solution is proven optimal (within the given tolerances), thereby providing a reliable benchmark for assessing the performance of the other methods. We note that, for all instances considered in this work, CPLEX Exact reaches a zero optimality gap within machine precision.
\item CPLEX Approx: This method also employs multi-threaded CPLEX, but introduces a time limit for the optimization procedure. The time constraint sacrifices the guarantee of finding an exact solution. We set the time limit to coincide with the runtime of SPLIT. Therefore, we can evaluate the practical advantage of SPLIT by assessing its ability to deliver solutions that surpass the quality of the ones returned by CPLEX Approx, within the same computational time. We note that, in order to prioritize the rapid identification of high-quality feasible solutions within CPLEX Approx, we use the \texttt{MIPEmphasis=5} option of CPLEX.
\end{itemize}
For all comparisons in this work, CPLEX was provided with the same number of computational cores used by SPLIT (i.e., equal to the number of subproblems $K$), for both the CPLEX Exact and CPLEX Approx settings, ensuring consistent computational resources across all methods.

Since the SPLIT method is heuristic in nature, in general we do not expect to obtain the exact solution $X_{\rm min}$, but rather a near-optimal one $X^*$. To address this point, we require a performance metric that quantifies the quality of the solutions produced by our heuristic approach. One way to estimate the proximity between the obtained solution and the exact one is through the approximation ratio, defined as
\begin{equation}\label{eq:alpha}
    \alpha = \frac{H(X^*)}{H(X_{\rm min})}\,,
\end{equation}
i.e. the ratio between the cost of the found solution $X^*$ and the optimal one. For relatively small instances of the problem, we can obtain an exact solution using CPLEX Exact. 

The second key performance metric we evaluate is the time-to-solution (TTS), defined as the time for the complete execution of the algorithm. In the case of SPLIT, the TTS indicates the timespan from  graph partitioning (included) to the achievement of convergence. For CPLEX Exact, TTS indicates the time required to find the optimal solution $X_{\rm min}$. We can then define the speed-up provided by SPLIT as the ratio
\begin{align}\label{eq:speedup}
    {\rm Speed-up} = \frac{{\rm TTS}_{\rm CPLEX-Exact}}{{\rm TTS}_{\rm SPLIT}}.
\end{align}
Speed-up and TTS are presented to illustrate the scaling of our method with the disclaimer that the values depend on the specific implementation and the optimization solver chosen for the subgraphs. Specifically, the performance could significantly improve if we switched from our Python implementation to a compiled language. For consistency, here we use the CPLEX solver as a subroutine to solve subproblems within the SPLIT method. An even larger speed-up could be achieved by using a different solver for the subproblems. Therefore, if the TTS achieved by our benchmark is lower than that of CPLEX, it suggests that our algorithm has the potential to perform even faster when implemented in other programming languages, which avoid the overhead introduced by the Python interpreter. 

\subsection{Implementation details}
We note that the TTS metric is also sensitive to the hardware used to run the optimization. For the calculations presented in this work, we used a single node of our proprietary \emph{davinci-1} cluster equipped with two Intel Cascade Lake Xeon Platinum 8260 @ 2.40\,GHz CPUs (24 cores each, 768 GB RAM). This configuration allows up to $K=48$ subproblems, each to be assigned to individual processors in the parallel section of SPLIT. Our implementation uses \emph{mpi4py} v3.1.4, a message-passing interface library~\cite{9439927}, to scatter the subproblems across the available processors/computational nodes. The corresponding MPI version is MPICH v3.3.2. The graph partitioning is performed by the functions of \emph{scikit-learn} v1.3.0 library~\cite{scikit-learn}. For what concerns the CPLEX solver~\cite{cplex2022v22}, we use the free license version v22.1, which is limited to a maximum of 1000 variables and 1000 constraints.

\section{Benchmark on the MaxCut problem}

\begin{figure}
    \centering
    \includegraphics[width=\linewidth]{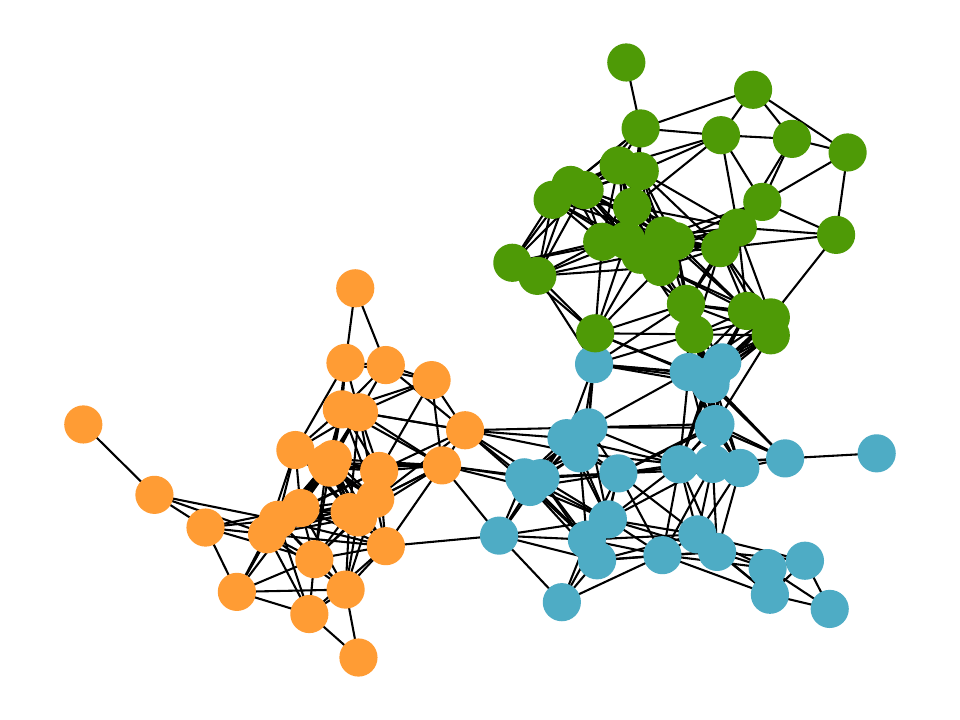}
    \caption{Graph with $N=90$ nodes partitioned by spectral clustering. For this graph, the SPLIT algorithm identifies the exact solution of the MaxCut problem. The exact solution takes TTS$_{\rm CPLEX-Exact} = 258 \; {\rm s}$, while SPLIT takes TTS$_{\rm SPLIT} = 3.3 \; {\rm s}$, reaching the exact solution after $N_{\rm iter} = 5$.}
    \label{fig:clusters}
\end{figure}

\subsection{The MaxCut problem}

As a first application, we tackle the MaxCut problem on a graph, which is represented by the cost function
\begin{align}
    H_{\rm MaxCut}(X) = -\sum_{(i,j) \in {E}} w_{ij} (x_i+x_j -2x_i x_j) \,,
\end{align}
which we seek to minimize~\cite{karp1972}. Here $x_i \in \{0,1\}$ are binary variables and $w_{ij}$ are real-valued numbers. We consider the unweighted case with $w_{ij}=1$, unless otherwise specified. To construct MaxCut problems with a fairly local structure, we generate the nodes of the graph by randomly placing $N$ points in the two-dimensional plane. The latter are sampled from distinct, isotropic Gaussian distributions, ensuring that they form spatially clustered groups ("blobs"). This is achieved by using the \texttt{sklearn.datasets.make\_blobs} function contained in the \texttt{scikit-learn} python package~\cite{sklearn_api}. In particular, the instances tested here are characterized by different values of standard deviations for the blobs, different seeds and different densities of the connections.
The global problem is defined as the MaxCut problem on the graph with nodes $\mathcal{N}= \{1, ..., N\}$. The edges $E$ are constructed based on a proximity criterion, where two nodes $i$ and $j$ are connected if their Euclidean distance falls below a certain threshold. 

In Fig. \ref{fig:clusters}, we show an example of a MaxCut instance with $N=90$ nodes, obtained by sampling three Gaussian blobs. The colors of the nodes indicate the different subgraphs obtained by applying the \emph{spectral clustering} algorithm (as implemented in \texttt{scikit-learn}) at the beginning of the SPLIT routine. The spectral clustering is able to correctly identify the original three-blob structure underlying the definition of the graph. We note that for this specific instance, the SPLIT method finds one of the exact solutions of the problem.

\begin{figure}
    \centering
    \includegraphics[width=\linewidth]{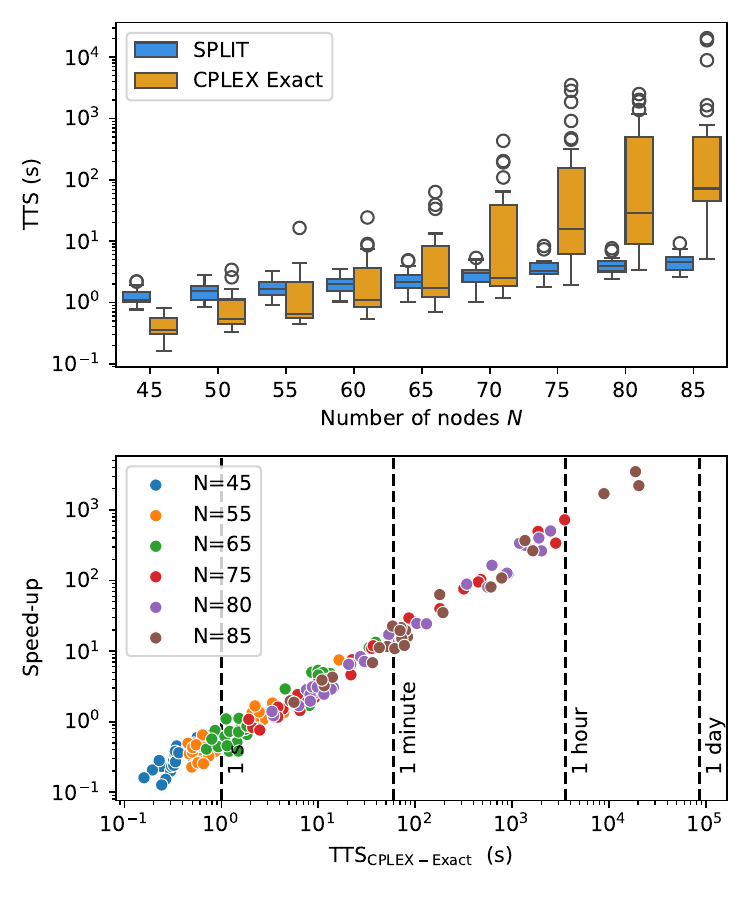}
    \caption{Upper panel: Boxplot showing the distribution of TTSs for SPLIT (blue boxes) and CPLEX Exact (orange boxes), on a semilog scale. Results for the MaxCut problem are displayed as a function of the number of graph nodes $N$  (30 instances for each problem size). The boxes represent the interquartile range. The horizontal lines within the boxes denote the median values. The whiskers span $1.5$ times the box range.
    Outliers located outside the range of the whiskers are displayed as empty circles.
    Lower panel: Log-log scatterplot of the speed-up of SPLIT as a function of the CPLEX Exact TTS for different instance sizes, highlighted by different colors.  }
    \label{fig:speedup}
\end{figure}

\subsection{SPLIT performance on random MaxCut instances}

\begin{figure}[ht]
    \centering
    \includegraphics[width=\linewidth]{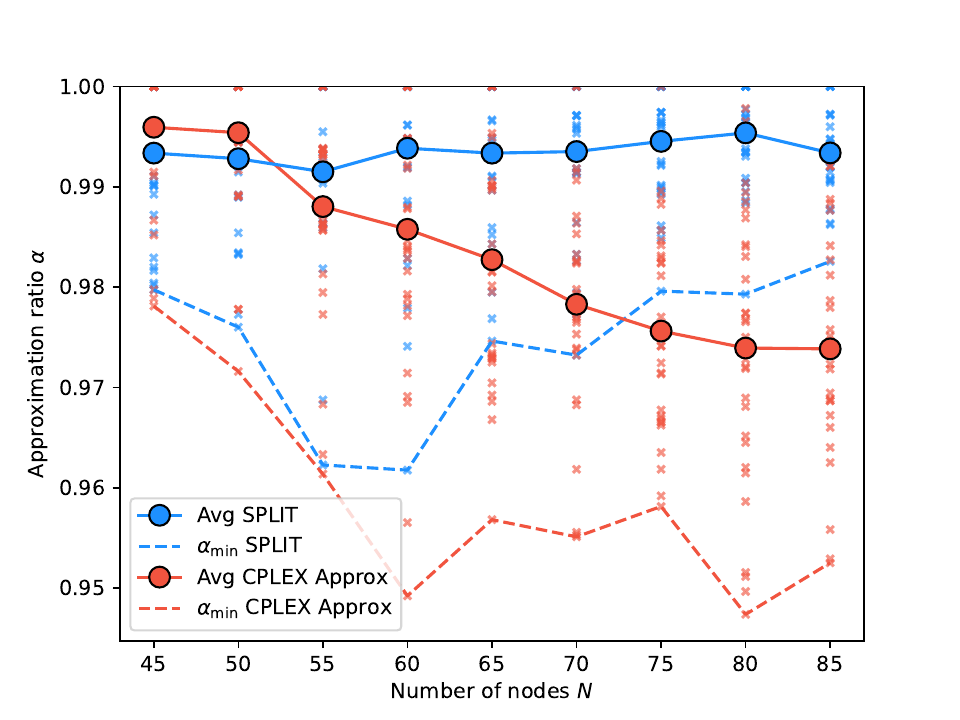}
    \caption{Approximation ratio~\eqref{eq:alpha} of the MaxCut solution with increasing $N$. The crosses represent the values of the approximation ratio achieved on the various instances. The average value of $\alpha$ for each graph size $N$ is shown by the dots connected by full lines, while the minimum approximation ratio observed among the different instances is indicated by the dashed lines. The blue markers indicate results obtained with SPLIT, while the red ones are CPLEX Approx results.}
    \label{fig:alpha_vs_cplex}
\end{figure}

We first test our heuristic approach in a regime in which the method is expected to perform well, namely instances generated starting from graphs with denser regions (blobs) and few interconnections between these regions. These graphs can be easily partitioned into clusters by removing a small number of edges. The latter are generated by sampling $4$ Gaussian blobs, changing the variance and seed of the Gaussians, as well as the distance threshold which controls the assignment of edges in the graph. Fig.~\ref{fig:speedup} illustrates a comparison of the TTSs of SPLIT and CPLEX Exact across multiple instances. While the SPLIT method is not guaranteed to find an exact solution, this comparison highlights its effectiveness in delivering near-optimal solutions within a significantly reduced computational time. Specifically, we generated $30$ instances and solved them using a partitioning in $K=4$ subproblems. 

The upper panel of Fig.~\ref{fig:speedup} displays a boxplot illustrating the TTS for both methods, plotted as a function of the number of graph nodes $N$. The blue boxplots correspond to the SPLIT method, while the orange boxplots represent the CPLEX Exact solver. The variation in box sizes reflects the distribution of solution times, highlighting differences in computational efficiency between the two approaches as the problem size increases. Specifically, the SPLIT method always provides a near-optimal solution within seconds, while an exact solution using CPLEX Exact takes up to several hours on the hardest instances, even for these relatively small graph sizes. The median speed-up demonstrates an improvement of several orders of magnitude with respect to CPLEX Exact as the size of the MaxCut problem increases, which may be crucial in time-critical applications. 

The lower panel of Fig.~\ref{fig:speedup} provides a direct analysis of the speed-up~\eqref{eq:speedup} for each individual instance. Specifically, each point corresponds to a different instance, with distinct colors representing different problem sizes. This visualization emphasizes how the computational advantage of SPLIT increases with increasing instance complexity, demonstrating its efficiency relative to the Exact CPLEX solver. In fact, SPLIT can find near-optimal solutions within seconds for this class of MaxCut problems across the graph sizes considered here, whereas CPLEX Exact can require minutes or even hours. The observed trends suggest that SPLIT achieves significant speed-ups, enabling the solution of large optimization problems in near real-time. This capability is crucial for industrial applications that demand solutions within strict time constraints.

Having assessed that SPLIT is able to find solutions faster than the exact CPLEX solver, we evaluate the quality of these solutions. We compare them to those found by CPLEX Approx, i.e. by fixing a time limitation for CPLEX when solving the full problem, equal to the time needed by SPLIT to solve the same instance. The result of this analysis is shown in Fig.~\ref{fig:alpha_vs_cplex}, where we plot the approximation ratio $\alpha$ for both SPLIT (blue curves) and CPLEX Approx (red curves). We report the average approximation ratio, as well as the minimum approximation ratio $\alpha_{\rm min}$ for the two methods, as a function of the number of graph nodes $N$. We observe that for small instances $\alpha$ is higher for CPLEX Approx. However, as $N$ is increased, the average quality of the CPLEX Approx solution degrades, while the SPLIT method displays values of $\alpha$ larger than 0.99 in the whole range considered here. At the same time, the $\alpha_{\rm min}$ of SPLIT becomes consistently larger than the one obtained with CPLEX Approx, signaling that the numerical lower bound on the SPLIT solution is better than the CPLEX Approx one.

\begin{figure}[t]
    \centering
    \includegraphics[width=\linewidth]{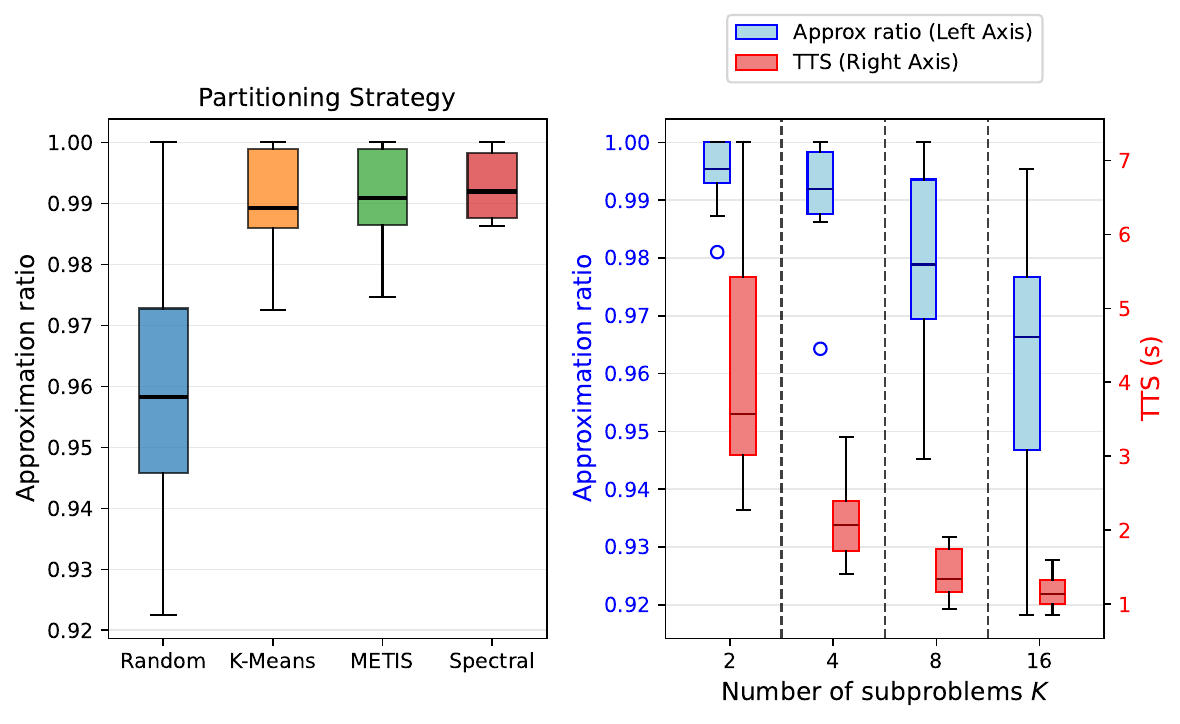}
    \caption{Left panel: approximation ratio achieved by SPLIT for the MaxCut problem on $18$ instances with $N=55$ nodes, using $K=4$ subproblems and different graph partitioning schemes (random partitioning, K-Means, METIS, spectral clustering). Right panel: approximation ratio and TTS of SPLIT as functions of $K$, using spectral clustering for graph partitioning. The same instances of the left panel are used. In both panels, boxplots follow the convention described in Fig.~\ref{fig:speedup}.}
    \label{fig:results_vs_clustering}
\end{figure}

To further investigate the role of the graph partitioning strategy and the choice of the number of subproblems $K$ within SPLIT, we perform an additional set of tests on fixed-size MaxCut instances. The left panel of Fig.~\ref{fig:results_vs_clustering} compares the performance of SPLIT under different partitioning schemes, including random partitioning, K-Means clustering~\cite{1056489}, METIS graph partitioning~\cite{doi:10.1137/S1064827595287997} and spectral clustering. We note that K-Means is applicable in this setting because the graph instances arise from spatial point distributions sampled from isotropic Gaussian blobs, thereby admitting a coordinate-based partitioning. The results show the importance of constructing weakly coupled subgraphs to enhance the effectiveness of the SPLIT iterations. A degradation in performance is observed if random partitioning is adopted. The other methods, which take into account the actual structure of the graph, have a similar performance among each other, with spectral clustering yielding the best overall results.

Sticking to spectral clustering, in the right panel of Fig.~\ref{fig:results_vs_clustering}, we analyze the dependence of both the approximation ratio and the TTS on the number of subproblems $K$. The average size of the subproblems decreases as $K$ is increased, leading to faster subproblem solutions and, thus, a reduced TTS. Naturally, this comes at the expense of a degradation in solution quality, due to the increased fragmentation of the original problem and the accumulation of inter-subproblems approximations. These results illustrate the trade-off between computational efficiency and solution accuracy inherent in the SPLIT framework. In practice, the choice of $K$ should be guided by both the available computational resources and the desired balance between speed and quality.

\begin{figure}[t]
    \centering
    \includegraphics[width=0.7\linewidth]{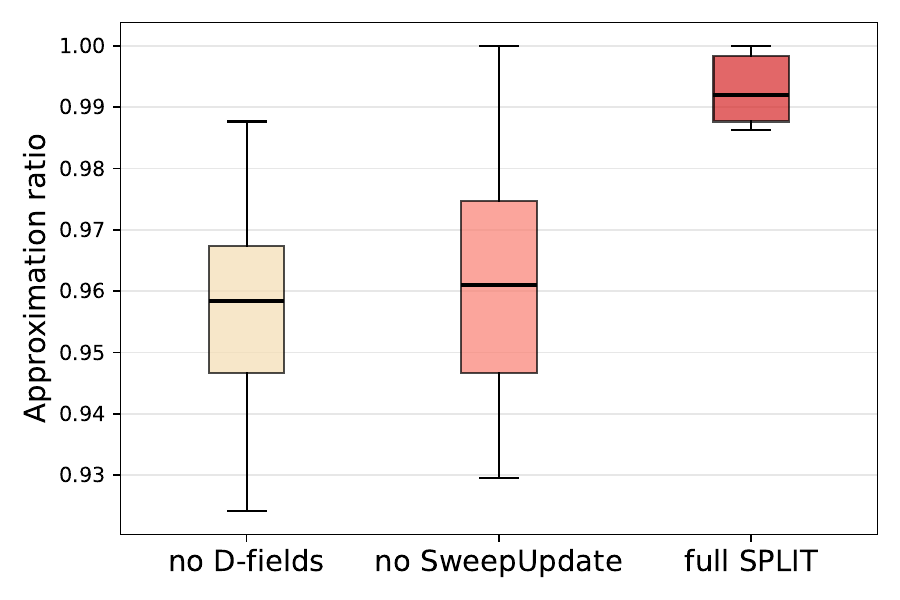}
    \caption{Approximation ratio achieved by different ablated variants of SPLIT for the MaxCut problem, evaluated on $18$ instances with $N=55$ nodes (using $K=4$ subproblems). The leftmost box corresponds to the case in which the $D$ fields are set to zero, the central box corresponds to SPLIT without the SweepUpdate routine, and the rightmost box refers to the full SPLIT pipeline.}
\label{fig:ablation}
\end{figure}

To complete our set of ablation tests, we assess the performance degradation of SPLIT in two limiting cases. First, we consider the case in which cross-interaction terms are neglected, i.e., the $D$ fields are fixed to zero. In this case, the algorithm terminates after a single iteration. Second, we consider the case in which the SweepUpdate routine is removed from the SPLIT pipeline. The results are shown in Fig.~\ref{fig:ablation} for the same fixed-size MaxCut instances used in Fig.~\ref{fig:results_vs_clustering}. Both ablated variants of SPLIT yield reduced performances, highlighting the crucial role of cross-interactions and SweepUpdate routine within the pipeline.

\subsection{Computational scaling of the algorithm}\label{sec:comp_scaling}

\begin{figure}[t]
    \centering
    \includegraphics[width=\linewidth]{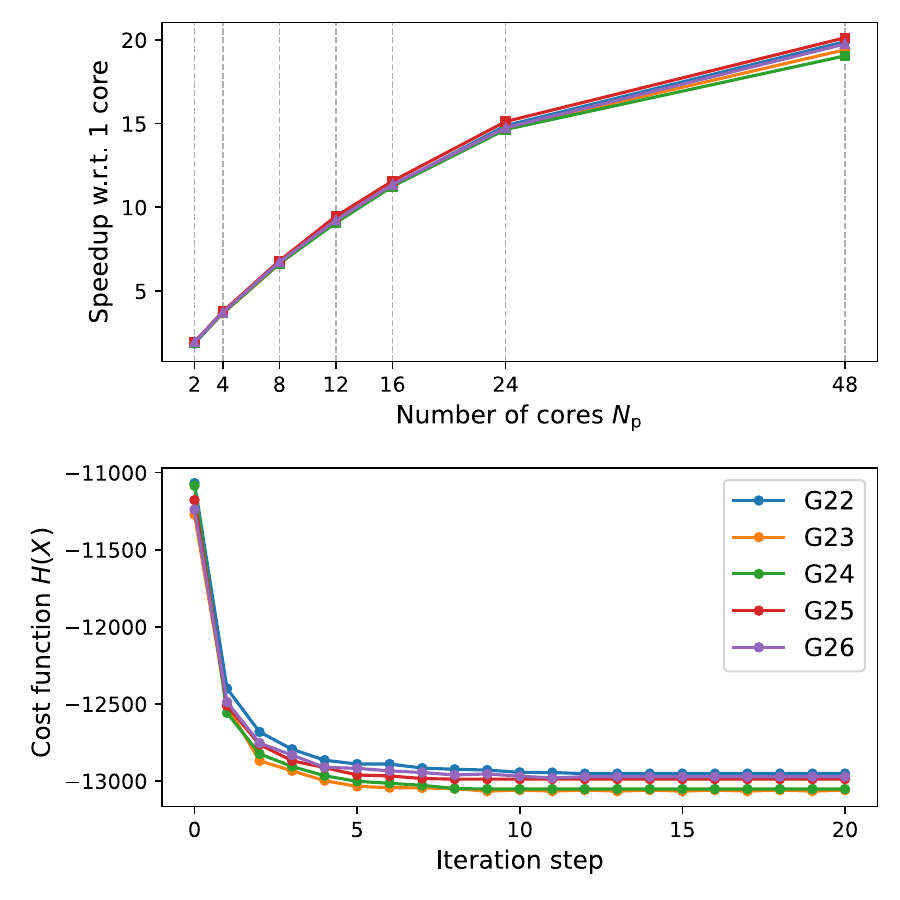}
    \caption{Upper panel: Speedup of parallel SPLIT relative to its single-core execution as a function of the number of parallel cores $N_p$. Results for Gset graphs G22-26 are reported, using a decomposition into $K=48$ subproblems. Excellent agreement with Amdahl’s law is observed, with an estimated parallel fraction $p = 0.97(1)$. 
    Lower panel: Evolution of the cost function value versus iteration step for G22–G26, showing empirical convergence.}
\label{fig:gset_convergence}
\end{figure}

The computational complexity of SPLIT depends linearly on the number of iterations. Each iteration comprises two stages: the (parallel) solution of the subproblems and the (sequential) Sweep Update. Since, in principle, subproblems may be heterogeneous in size and difficulty, let us denote by $t_{\rm sub}$ the worst-case subproblem solve time for the chosen solver. In the case of our numerical benchmarks, since the branch-and-bound algorithm of CPLEX is used as subproblem solver, we can estimate $t_{\rm sub}\sim 2^{\max\limits_{k} N_k}$. Let us then indicate by $t_{\rm sweep}$ the time required by the Sweep Update routine, which in our case has a complexity that scales linearly with the number of edges, i.e. $O(|E|)$. Assuming that $N_p$ computational resources are available to tackle the subproblems in parallel, e.g. $N_p$ CPU cores, we can express the overall worst-case runtime of SPLIT as $t_{\rm par}+N_{\rm iter}( t_{\rm sub} \lceil \frac{K}{N_p}\rceil + t_{\rm sweep})$, where $t_{\rm par}$ is the time required for graph partitioning. Therefore, the availability of $N_p$ parallel resources can help reducing the burden of the subproblem solution step, which is the heaviest step of the pipeline in this context.

To showcase the scalability of SPLIT with the number of parallel resources, we perform a numerical experiment using five medium-scale graphs from the Gset benchmark collection of MaxCut instances, a widely used set of test instances in combinatorial optimization~\cite{doi:10.1137/S1052623497328987,MaxCutBenchmark}. The Gset dataset includes a diverse set of graphs with varying structures, densities, and non-uniform edge weights ($w_{ij}=\pm 1$), making it a suitable benchmark for assessing the effectiveness of optimization algorithms. Here we consider the graphs G22-G26, each consisting of 2000 nodes, which we partition into $K=48$ subproblems. This implies that, on average, each subproblem has a size of approximately $40$ nodes, which can be solved with the CPLEX subproblem solver within few seconds of runtime. We perform a fixed number of $N_{\rm iter}=250$ steps, without stopping in case convergence is achieved early. Most importantly, each calculation is performed parallelizing the subproblem solution step over different number of cores, $N_p\in\{1,2,4,8,16,24,48\}$. In the upper panel of Fig.~\ref{fig:gset_convergence}, we report the speedup achieved with $N_p>1$ cores compared to the single-core calculation, as a function of the number of cores $N_p$. The data are well described by Amdahl’s law~\cite{10.1145/1465482.1465560}, yielding a fitted parallel fraction of $p = 0.97(1)$. This demonstrates how SPLIT can largely benefit from parallel execution. For completeness in the lower panel of Fig.~\ref{fig:gset_convergence} we report the value of the cost function as a function of the iterations, for the various instances. The cost function value rapidly approaches a plateau in all cases considered, giving empirical indication of convergence of the iterative procedure. However, no formal convergence proof is currently available.

\subsection{Performance on large-scale Gset graphs}
\begin{table}[h]
    \caption{Comparison of SPLIT cost and the best known cost for different large-scale instances of Gset (see.~\cite{Shylo2015,zick2025performancereportheuristicalgorithm}). \label{tab:comparison}}
    \centering
    \begin{tabular}{lccccc} 
\hline
        \textbf{Instance} & \textbf{Nodes} & \textbf{SPLIT Cost} & \textbf{Best Cost} & \textbf{TTS (s)} & \textbf{$\alpha$}\\
\hline
G65 & 8000 & 5482 & 5562 & 1198 & 0.986 \\
G66 & 9000 & 6282 & 6364 & 1303 & 0.987\\
G67 & 10000 & 6866 & 6950 & 2334 & 0.988\\
G72 & 10000 & 6918 & 7008 & 2550 & 0.987 \\
G77 & 14000 & 9874 & 9940 & 5322 & 0.993\\
G81 & 20000 & 13986 & 14060 & 15049 & 0.995\\
\hline
    \end{tabular}    
\end{table}

As part of our study on the MaxCut problem, we apply the SPLIT algorithm to a selected set of large-scale Gset instances with sizes up to 20,000 nodes to further assess its effectiveness on increasingly large and demanding instances. These results are well-beyond the capabilities of Exact CPLEX. The results are summarized in Table~\ref{tab:comparison} and are obtained using $K=48$ and fixing a timelimit of 10 seconds for the subproblem solver at each iteration. The maximum number of performed iterations is 500. Although our matheuristic method does not beat specialized heuristics on this specific problem~\cite{BENLIC20131162, Shylo2015, Ma2017, zick2025performancereportheuristicalgorithm}, these results demonstrate the ability of the SPLIT algorithm to achieve high-quality cuts across different graph instances within relatively short execution times. For instance, compared to the results reported in Ref.~\cite{BENLIC20131162}, SPLIT achieves cuts that are, on average, within $1\%$ of the reported values, while requiring between $25\%$ and $80\%$ less computational time.

\section{Antenna placement problem}

\subsection{The Antenna Placement Problem}

\begin{figure}
    \centering
    \includegraphics[width=\linewidth]{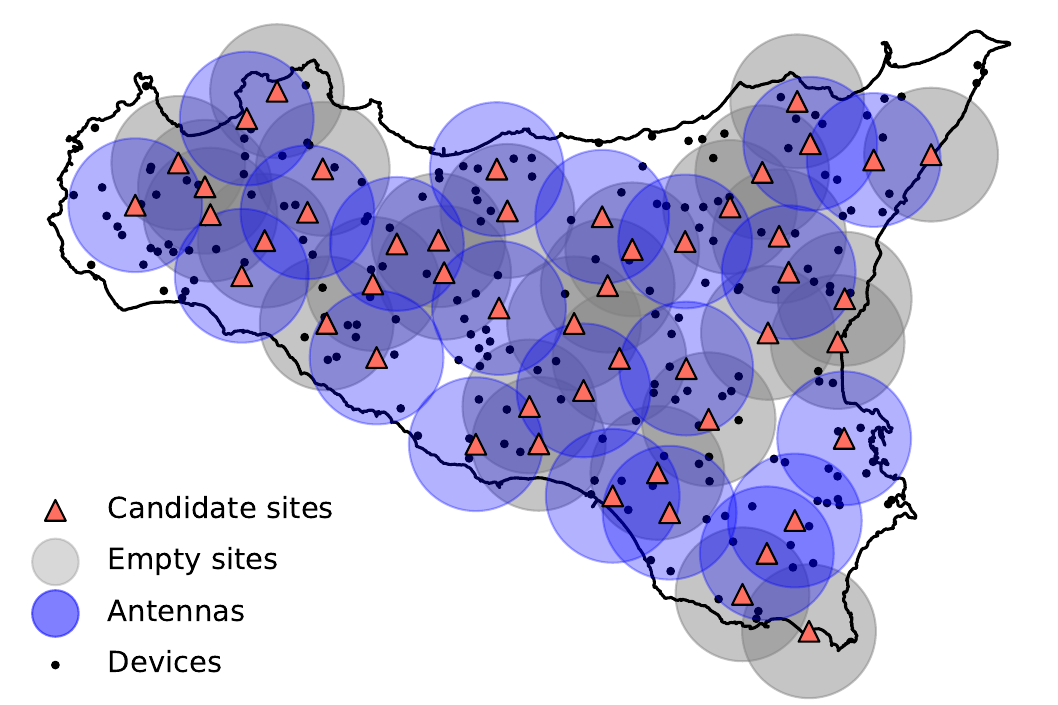}
    \caption{Example of a randomly generated APP instance on an Italian region, specifically on Sicily, with $N=50$ sites and $V=25$ antennas. The orange triangles represent the candidate sites for antennas, while the black dots are devices that need to be optimally covered. The circles illustrate the area covered by a (potential) antenna. Blue circles are those assigned to an antenna, while gray circles are empty sites. The result shown here is the one obtained by SPLIT, which coincides with the optimal solution in this case.}
\label{fig:example_sicily}
\end{figure}

To demonstrate the flexibility and effectiveness of SPLIT, we apply it to the Antenna Placement Problem (APP), a prototypical industrial challenge previously described in Ref.~\cite{vandelli2024evaluating}. This problem involves the identification of a large number of potential deployment sites for antennas, while only a limited number of antennas are available. The goal is to optimally distribute the antennas across the selected sites, in order to maximize signal coverage, while minimizing interference between antenna signals. 
In this specific example, signal coverage refers to reaching a set of devices distributed on the territory. This application has relevant implications for emergency response and search \& rescue operations~\cite{vandelli2024evaluating}. To draw a clear connection with practical applications, we consider realistic antenna distributions generated on the 20 regions of Italy as instances of the problem. An illustrative example is given in Fig.~\ref{fig:example_sicily}. Similarly to several other industrial problems, and contrary to the vanilla MaxCut, the APP is highly unstructured, has continuous weights and lacks symmetries. Nevertheless, it exhibits key features, such as locality of the cost function as a consequence of the geographical arrangement of the antennas, and a simple constraint on the number of occupied sites.

The cost function of the APP discussed here reads
\begin{align}
    H_{\rm APP}(X) &= \sum_{i < j} O_{ij} x_i x_j - \frac{1}{4} \sum_{i=1}^N A_i x_i \\
    &{\rm s.t} \quad \sum_{i=1}^N x_i = V
\end{align}
where each potential site $i=1,...,N$ is described by a single binary variable $x_i$. The coefficients $O_{ij}$ represent the potential overlap area between pairs of sites $(i,j)$, $A_i$ is the area covered by the signal of an antenna located at site $i$, and $V$ is the total number of available antennas~\cite{vandelli2024evaluating}. The effective areas are calculated by counting the number of devices located within the coverage range of the antennas. In the example of Fig.~\ref{fig:example_sicily}, we take the antenna coverage radius to be $15$ km. When studying the scaling behaviour as a function of the problem size $N$, we scale the radius by $1/\sqrt{N}$ to ensure a reasonable overlap is maintained across the whole range of sizes considered.
For this problem, we enforce the cardinality constraint exactly using the scheme in Eq.~\eqref{eq:cardinality_const}, ensuring that the near-optimal solutions found by SPLIT always satisfy it.

\subsection{Scaling \emph{vs.} problem size}

\begin{figure}[t]
    \centering
    \includegraphics[width=\linewidth]{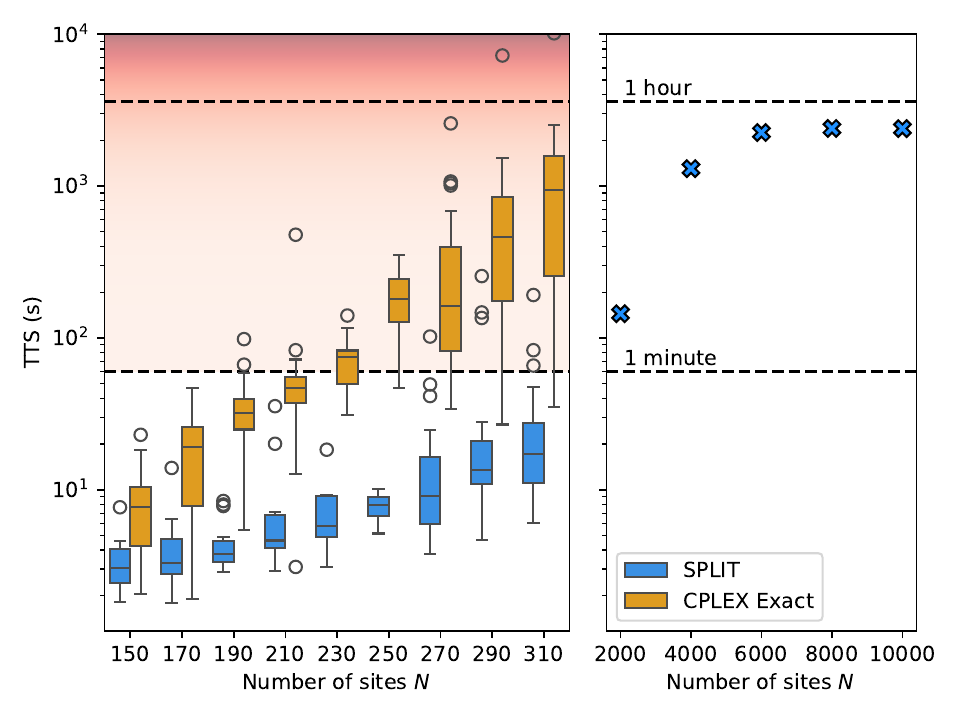}
    \caption{Left panel: Boxplot showing the distribution of TTSs for SPLIT with $K=6$ (blue boxes) and CPLEX Exact (orange boxes). Statistics is collected over 20 instances of the APP for each number of sites $N$. The dashed lines denote $1$ minute and $1$ hour on the TTS axis. Right panel: Semilog scatterplot displaying the TTSs for the largest APP instances considered here, up to $N=10,000$ sites. Here, the number of SPLIT subproblems is increased as the problem size increases, according to $K = \lfloor N/50 \rfloor$. }
    \label{fig:time_antennas}
\end{figure}

In this section, we investigate the performance of SPLIT for the APP across different problem sizes. 
We start by comparing the runtimes of SPLIT (blue boxes) and CPLEX Exact (orange boxes) in Fig.\ref{fig:time_antennas}. In the left panel, we report the results for the smallest sizes considered here, $N\leq 230$. For these calculations, we set the number of subproblems to $K=6$. The red color gradient in the figure highlights the increasing computational time, with a threshold of one hour for realistic applications. CPLEX Exact exceeds this threshold as early as problem size $N=290$, for certain instances. TTS$_{\rm SPLIT}$, instead, remains below one minute for most of the instances considered here. Although SPLIT significantly reduces computational time compared to CPLEX Exact, its scaling remains exponential when the number of subproblems is fixed. Indeed, for a constant value of $K$, the size of each subproblem $H_k$ increases as $N$ grows. This limits the maximum problem size reachable with a fixed $K$. 

In order to push the method to much larger problem sizes, we can introduce a rule to increase $K$ as $N$ is increased, in a way which ensures that the subproblem size remains sufficiently small and that the computational effort is distributed across a proportionally larger number of subproblems.
Consequently, in our calculations at larger $N$, $K$ is chosen according to the relation \( K = \lfloor N/50 \rfloor \). 
This approach not only allows for solving much larger problem sizes on standard computing resources, but also mimics scenarios where problem sizes must be significantly reduced to fit the hardware constraints of specialized devices, such as those required in edge computing (e.g. limited power and memory) or near-term quantum computing (restricted qubit count on each QPU). In the right panel of Fig. \ref{fig:time_antennas}, we demonstrate that this strategy can be applied to reach problem sizes with thousands of sites, remaining within reasonable time constraints for realistic applications. To support this statement, we consider a single instance for different values of $N$ between $2000$ and $10,000$, in which antenna sites are distributed across the entire Italian territory. In this case, the time behavior of SPLIT is approximately linear as we run the calculations on a single computational node and, consequently, the number of subproblems assigned to each processor grows linearly with $N$.

\begin{figure}[t]
    \centering
    \includegraphics[width=0.8\linewidth]{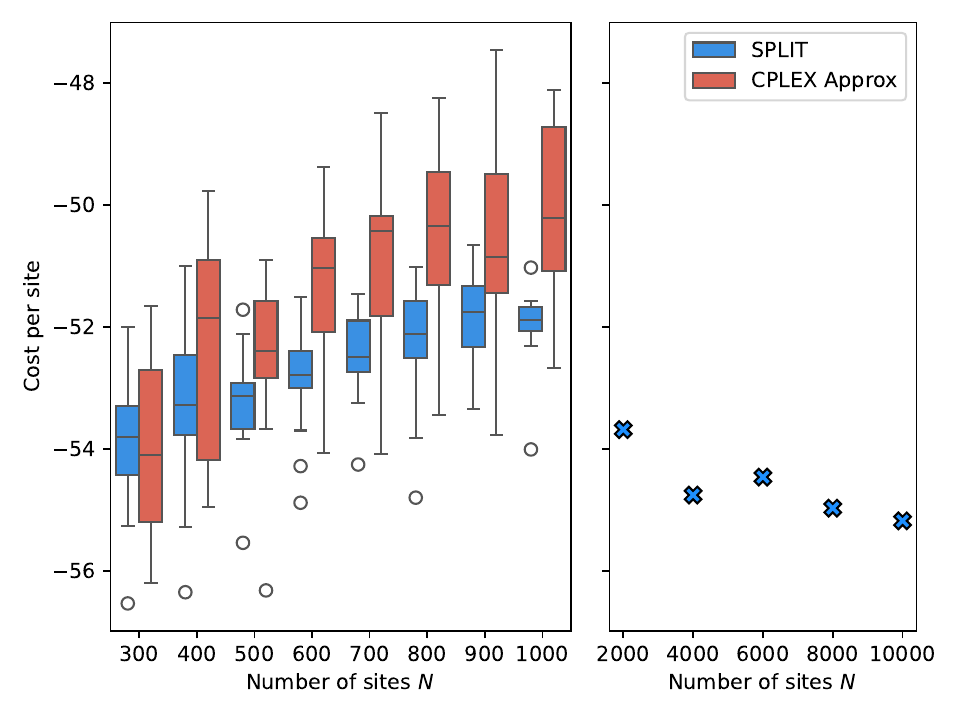}
    \caption{
    Left panel: Boxplot showing the distribution of the cost per site  obtained by SPLIT and CPLEX Approx for the APP over 20 instances. The costs are plotted as a function of the number of sites $N$. Right panel: Scatterplot of the costs per site achieved by SPLIT for the largest APP instances considered here, up to $N=10,000$ sites.
    In both panels, the number of SPLIT subproblems is increased as the problem size increases, according to $K = \lfloor N/50 \rfloor$.}
    \label{fig:cost_fixed_tts}
\end{figure}

After discussing the time performance of SPLIT on the APP, we compare the solution quality of SPLIT and CPLEX Approx. The results are shown in Fig.\ref{fig:cost_fixed_tts}, where we report the cost function per site $H(X^*)/N$, for increasing values of $N$. The number of subgraphs is once more chosen according to the relation \( K = \lfloor N/50 \rfloor \). Since the antennas are distributed roughly homogeneously, the median cost varies only marginally across all the instances as the size increases. In the left panel of Fig. \ref{fig:cost_fixed_tts}, we show results for $N\leq 1000$. In this range, we can draw a comparison of the costs obtained using SPLIT (blue boxes) and CPLEX Approx (red boxes). For the smallest problem sizes considered here with $N = 300$ and $N=400$, SPLIT and CPLEX Approx perform similarly. However, as $N$ increases, the cost of CPLEX Approx degrades compared to SPLIT, displaying a finite cost gap between the two methods. In the right panel of Fig. \ref{fig:cost_fixed_tts}, we report the cost of the $N\geq 2000$ instances of Fig. \ref{fig:time_antennas}, to demonstrate that SPLIT still retrieves solutions with a reasonable cost even for industrially relevant problem scales. 

\section{Conclusions}

In this work, we introduced SPLIT, a heuristic approach for decomposing large-scale quadratic programs---potentially involving thousands of continuous, integer, or mixed variables---into smaller, more manageable subproblems. In our method, the latter are solved in parallel, drastically reducing the computational time compared to solving the original problem as a whole. Unlike previous decomposition techniques that treat subproblems as independent, SPLIT makes use of an iterative scheme to approximately account for inter-dependencies between variables across different subproblems. Our method can incorporate certain types of constraints exactly, offering adaptability for handling realistic industrial problems.

SPLIT is designed to leverage HPC resources, efficiently distributing subproblems across multiple computing devices, which can include conventional CPUs or accelerators such as GPUs, FPGAs, and QPUs. The problem decomposition implemented in SPLIT reduces resource demands without requiring a substantial modification of the optimization pipeline. In fact, the SPLIT framework is solver-agnostic and can seamlessly integrate a variety of subproblem solvers, from commercial solutions based on branch-and-bound methods, to emerging quantum and quantum-inspired approaches, as demonstrated in a follow-up work that leverages a quantum optimization algorithm to solve the subproblems~\cite{vandelli2025constraintpreservingquantumalgorithmmultifrequency}. 

To show the effectiveness of SPLIT, we compared its performance against the commercial CPLEX solver for two distinct use cases: the MaxCut problem, a well-established benchmark in unconstrained combinatorial optimization, and the Antenna Placement Problem, a simplified yet representative industrial application that features a capacity constraint. Our results highlight the computational advantages of our method with respect to exact solvers, especially for large problem sizes with hundreds or even thousands of variables. We achieve reductions in computational time by orders of magnitude compared to the direct application of the exact CPLEX solver run on multiple cores. This improvement is particularly relevant for large-scale problems in which conventional solvers face challenges with scalability. For near real-time applications, which require rapid delivery of a solution, we show that within the same fixed runtime, SPLIT provides higher quality results compared to CPLEX. These results establish the advantage of using SPLIT for large-scale optimization problems. 

Future directions include evaluating the performance of SPLIT on problems with non-binary variables, particularly mixed-integer problems, as well as exploring techniques to effectively encode additional types of hard constraints within the framework. Additionally, improvements to the algorithm, such as treating nodes with external connections differently than internal nodes, or dynamically reassigning variables to the subproblems~\cite{vandelli2025constraintpreservingquantumalgorithmmultifrequency}, could enhance the overall efficiency and solution quality.

\section*{Statements and Declarations}
The authors declare no competing financial interests.

\section*{Data Availability Statement}
The data supporting the results of this study are available from the authors upon reasonable request.

\bibliographystyle{unsrturl}
\bibliography{sn-bibliography}

\end{document}